\documentclass[10pt]{article}
\usepackage{graphicx}
\usepackage{amsmath}
\usepackage{amssymb}
\usepackage{caption2}
\begin{document}
\begin{center}
{\large {\bf \sc{  Masses and  decay constants of the heavy tensor  mesons with  QCD sum rules }}} \\[2mm]
Zhi-Gang Wang \footnote{E-mail,zgwang@aliyun.com.  }, Zun-Yan Di      \\
 Department of Physics, North China Electric Power University,
Baoding 071003, P. R. China
\end{center}

\begin{abstract}
In this article, we calculate the contributions of the vacuum condensates up to dimension-6 in the operator product expansion,  study the masses and decay constants of the heavy tensor mesons  $D_2^*(2460)$, $D_{s2}^*(2573)$, $B_2^*(5747)$, $B_{s2}^*(5840)$  using the QCD sum rules.
The predicted masses   are in excellent agreement with the experimental data, while the ratios of the decay constants $\frac{f_{D_{s2}^*}}{f_{D_{2}^*}}\approx\frac{f_{B_{s2}^*}}{f_{B_{2}^*}}\approx\frac{f_{D_{s}}}{f_{D}}\mid_{\rm exp}$, where the exp denotes the experimental value.
\end{abstract}



\section{Introduction}

In recent years, the Babar, Belle, CLEO, CDF, D0, LHCb  and BESIII collaborations have observed  (or
confirmed) many  charmonium-like states and revitalized the interest in the spectroscopy
of the charmonium states \cite{Swanson2006}. There are also some progresses in the spectroscopy of the conventional heavy mesons,
several excited states have been observed, such as the $D_{s1}^*(2700)$, $D_{sJ}^*(2860)$, $D_{sJ}(3040)$, $D_J(2580)$, $D_J^*(2650)$, $D_J(2740)$, $D^*_J(2760)$,  $D_J(3000)$, $D_J^*(3000)$, $B_1(5721)$, $B_2^*(5747)$, $B_{s1}(5830)$, $B_{s2}^*(5840)$, $B(5970)$, etc \cite{PDG}.
In 2007, the D0 collaboration firstly observed the $B_1(5721)^0$ and $B_2(5747)^0$   \cite{D0-2007},
later the CDF   collaboration confirmed  them  \cite{CDF-2008}.
Also in 2007, the CDF  collaboration observed the $B_{s1}(5830)$ and $B_{s2}^*(5840)$ \cite{CDF-Bs-2007}.
The D0 collaboration confirmed the  $B_{s2}^*(5840)$ \cite{D0-Bs-2007}.
In 2012, the LHCb collaboration updated the masses $M_{B_{s1}}=(5828.40\pm 0.04\pm 0.04\pm 0.41)\,\rm{ MeV}$ and $M_{B_{s2}^*}=(5839.99\pm 0.05\pm 0.11\pm 0.17)\,\rm{MeV}$ \cite{LHCb-2012}.
The heavy-light  mesons can be  classified in doublets according to the total
angular momentum of the light antiquark ${\vec s}_\ell$,
${\vec s}_\ell= {\vec s}_{\bar q}+{\vec L} $, where the ${\vec
s}_{\bar q}$ and ${\vec L}$ are the spin and orbital angular momentum of the light antiquark,  respectively. Now the $J^P_{s_\ell}=(0^+,1^+)_{\frac{1}{2}}$ doublets $(D_0^*(2400),\,D_1^{\prime}(2430))$, $(D_{s0}^*(2317),\,D_{s1}^{\prime}(2460))$, the  $J^P_{s_\ell}=(1^+,2^+)_{\frac{3}{2}}$ doublets   $(D_1(2420),\,D_2^*(2460))$,  $(D_{s1}(2536),\,D_{s2}^*(2573))$,
 $(B_1(5721),\,B_2^*(5747))$, $(B_{s1}(5830),\,B_{s2}^*(5840))$ are complete \cite{PDG}.

The  QCD sum rules is a powerful nonperturbative theoretical tool in studying the
ground state hadrons \cite{SVZ79,Reinders85,NarisonBook,ColangeloReview}. In the
QCD sum rules, the operator product expansion is used to expand the
time-ordered currents into a series of quark and gluon condensates
which parameterize the long distance properties. We can obtain copious information on
the hadronic parameters at the phenomenological side by taking
the quark-hadron duality
\cite{SVZ79,Reinders85,NarisonBook,ColangeloReview}. There have been many works on the
$J^P_{s_\ell}=(0^-,1^-)_{\frac{1}{2}}$ doublets with the full QCD sum rules \cite{NarisonBook,ColangeloReview} (or Ref.\cite{HM-QCDSR} for recent works), while the works on the $J^P_{s_\ell}=(0^+,1^+)_{\frac{1}{2}}$  and $(1^+,2^+)_{\frac{3}{2}}$ doublets  are few \cite{HM-12-P,HM-32-P,HM-32-P-2}. For the works on the QCD sum rules combined with the heavy quark effective theory, one can consult Refs.\cite{Neubert-1994,HMEFT-SR}.
In Ref.\cite{HM-32-P}, H. Sundu et al  study the masses and decay constants of the  tensor mesons $D_2^*(2460)$ and $D_{s2}^*(2573)$  with the QCD sum rules by taking into account the perturbative terms and the mixed condensates in the operator product expansion, the contributions of the gluon condensate, three-gluon condensate and four-quark condensate are neglected. Neglecting the vacuum condensates of dimension-4 and 6 impairs  the predictive ability.

In this article, we calculate the contributions of the  vacuum condensates up to dimension-6 in the operator product expansion consistently,  study the masses and decay constants of the heavy tensor mesons $D_2^*(2460)$, $D_{s2}^*(2573)$, $B_2^*(5747)$, $B_{s2}^*(5840)$ with the QCD sum rules.

The article is arranged as follows:  we derive the QCD sum rules for
the masses and decay constants of the heavy tensor  mesons  in Sect.2;
in Sect.3, we present the numerical results and discussions; and Sect.4 is reserved for our
conclusions.

\section{QCD sum rules for  the heavy tensor  mesons }
In the following, we write down  the two-point correlation functions
$\Pi_{\mu\nu\alpha\beta}(p)$  in the QCD sum rules,
\begin{eqnarray}
\Pi_{\mu\nu\alpha\beta}(p)&=&i\int d^4x e^{ip \cdot (x-y)} \langle
0|T\left\{J_{\mu\nu}(x)J_{\alpha\beta}^{\dagger}(y)\right\}|0\rangle\mid_{y=0} \, , \\
J_{\mu\nu}(x)&=&i\overline{Q}(x)\left( \gamma_\mu\stackrel{\leftrightarrow}{\partial}_\nu +\gamma_\nu\stackrel{\leftrightarrow}{\partial}_\mu-\frac{2}{3}\widetilde{g}_{\mu\nu} \stackrel{\leftrightarrow}{\!\not\!\partial}\right) q(x) \, , \\
\stackrel{\leftrightarrow}{\partial}_\mu&=&\stackrel{\rightarrow}{\partial}_\mu-\stackrel{\leftarrow}{\partial}_\mu \, ,\nonumber\\
\widetilde{g}_{\mu\nu}&=&g_{\mu\nu}-\frac{p_{\mu}p_{\nu}}{p^2} \, ,\nonumber
\end{eqnarray}
where $Q=c,b$ and $q=u,d,s$, the tensor currents $J_{\mu\nu}(x)$  interpolate the heavy tensor mesons $D_2^*(2460)$, $D_{s2}^*(2573)$, $B_2^*(5747)$ and $B_{s2}^*(5840)$, respectively. In Ref.\cite{Aliev1982},  Aliev and Shifman take the tensor currents $\eta^1_{\mu\nu}(x)$,
\begin{eqnarray}
\eta^1_{\mu\nu}(x)&=&\frac{1}{2}i\overline{q}(x)\left( \gamma_\mu\stackrel{\leftrightarrow}{D}_\nu +\gamma_\nu\stackrel{\leftrightarrow}{D}_\mu\right) q(x)\, ,
\end{eqnarray}
with $D_\mu=\partial_\mu-ig_sG_\mu$ and the $G_\mu$ is the gluon field,  to study the light tensor mesons.  Later, Reinders, Yazaki, Rubinstein take the tensor currents $\eta_{\mu\nu}^2(x)$,
\begin{eqnarray}
\eta^2_{\mu\nu}(x)&=&i\overline{q}(x)\left( \gamma_\mu\stackrel{\leftrightarrow}{\partial}_\nu +\gamma_\nu\stackrel{\leftrightarrow}{\partial}_\mu-\frac{2}{3}\widetilde{g}_{\mu\nu} \stackrel{\leftrightarrow}{\!\not\!\partial}\right) q(x) \, ,
\end{eqnarray}
to study the light tensor mesons \cite{Reinders1982}. In Ref.\cite{Bagan1988}, Bagan and Narison restudy the light tensor mesons with the tensor currents $\eta^1_{\mu\nu}(x)$. For recent works on the light tensor mesons with the QCD sum rules, see Ref.\cite{Light-tensor}. We have two choice, i.e. we can choose either the partial  derivative $\partial_\mu$ or the covariant derivative $D_\mu$ in constructing  the interpolating  currents. The currents $\eta^1_{\mu\nu}(x)$ with the covariant derivative $D_\mu$ are gauge invariant, but blur the physical interpretation of the $\stackrel{\leftrightarrow}{D}_\mu$ as the angular momentum; on the other hand,  the currents $\eta^2_{\mu\nu}(x)$ with the partial derivative $\partial_\mu$ are not gauge invariant, but manifest the physical interpretation of the $\stackrel{\leftrightarrow}{\partial}_\mu$ as the angular momentum. In this article, we will present the results come from the currents with both the partial derivative and the covariant derivative.

We can insert  a complete set of intermediate hadronic states with
the same quantum numbers as the current operators $J_{\mu\nu}(x)$ into the
correlation functions $\Pi_{\mu\nu\alpha\beta}(p)$  to obtain the hadronic representation
\cite{SVZ79,Reinders85}. After isolating the ground state
contributions from the heavy tensor mesons, we get the following result,
\begin{eqnarray}
\Pi_{\mu\nu\alpha\beta}(p)&=&\frac{f_{T}^2M_{T}^4}{M_{T}^2-p^2}{\rm P}_{\mu\nu\alpha\beta} +\cdots\,  ,\nonumber\\
&=&\Pi(p^2){\rm P}_{\mu\nu\alpha\beta}\,  ,
\end{eqnarray}
where the  decay constants $f_{T}$ are defined by
\begin{eqnarray}
\langle 0|J_{\mu\nu}(0)|T(p)\rangle&=&f_{T}M^2_{T}\varepsilon_{\mu\nu}  \, ,
\end{eqnarray}
the $\varepsilon_{\mu\nu}$ are the polarization vectors of the tensor mesons  with the following properties \cite{JJZhu},
\begin{eqnarray}
{\rm P}_{\mu\nu\alpha\beta}&=&\sum_{\lambda} \varepsilon^*_{\mu\nu}(\lambda,p)\varepsilon_{\alpha\beta}(\lambda,p)=\frac{\widetilde{g}_{\mu\alpha}\widetilde{g}_{\nu\beta}+\widetilde{g}_{\mu\beta}\widetilde{g}_{\nu\alpha}}{2}
-\frac{\widetilde{g}_{\mu\nu}\widetilde{g}_{\alpha\beta}}{3} \, , \nonumber\\
p^{\mu}{\rm P}_{\mu\nu\alpha\beta}&=&\widetilde{g}^{\mu\nu}{\rm P}_{\mu\nu\alpha\beta}=\widetilde{g}^{\alpha\beta}{\rm P}_{\mu\nu\alpha\beta}=g^{\mu\nu}{\rm P}_{\mu\nu\alpha\beta}=g^{\alpha\beta}{\rm P}_{\mu\nu\alpha\beta}=0\, ,\nonumber\\
5&=&{\rm P}_{\mu\nu\alpha\beta}{\rm P}^{\mu\nu\alpha\beta}\, .
\end{eqnarray}

Now, we briefly outline  the operator product
expansion for the correlation functions $\Pi_{\mu\nu\alpha\beta}(p)$  in perturbative
QCD.  We contract the quark fields in the correlation functions
$\Pi_{\mu\nu\alpha\beta}(p)$ with Wick theorem firstly,
\begin{eqnarray}
\Pi(p^2)&=&\frac{1}{5}{\rm P}^{\mu\nu\alpha\beta}\Pi_{\mu\nu\alpha\beta}(p)\nonumber\\
&=&-\frac{i}{5}{\rm P}^{\mu\nu\alpha\beta}\int d^4x e^{ip \cdot (x-y)}   Tr\left\{\Gamma_{\mu\nu}S_{ij}(x-y)\Gamma_{\alpha\beta} S^Q_{ji}(y-x) \right\}\mid_{y=0}\, ,
\end{eqnarray}
where
\begin{eqnarray}
\Gamma_{\mu\nu}&=&i\left( \gamma_\mu\stackrel{\leftrightarrow}{\frac{\partial}{\partial x^\nu}} +\gamma_\nu\stackrel{\leftrightarrow}{\frac{\partial}{\partial x^\mu}}-\frac{2}{3}\widetilde{g}_{\mu\nu} \gamma^\tau\stackrel{\leftrightarrow}{\frac{\partial}{\partial x^\tau}}\right)\, ,\nonumber\\
\Gamma_{\alpha\beta}&=&i\left( \gamma_\alpha\stackrel{\leftrightarrow}{\frac{\partial}{\partial y^\beta}} +\gamma_\beta\stackrel{\leftrightarrow}{\frac{\partial}{\partial y^\alpha}}-\frac{2}{3}\widetilde{g}_{\alpha\beta} \gamma^\tau\stackrel{\leftrightarrow}{\frac{\partial}{\partial y^\tau}}\right)\, ,
\end{eqnarray}
\begin{eqnarray}
S_{ij}(x)&=& \frac{i\delta_{ij}\!\not\!{x}}{ 2\pi^2x^4}
-\frac{\delta_{ij}m_q}{4\pi^2x^2}-\frac{\delta_{ij}}{12}\langle
\bar{q}q\rangle +\frac{i\delta_{ij}\!\not\!{x}m_q
\langle\bar{q}q\rangle}{48}-\frac{\delta_{ij}x^2\langle \bar{q}g_s\sigma Gq\rangle}{192}+\frac{i\delta_{ij}x^2\!\not\!{x} m_q\langle \bar{q}g_s\sigma
 Gq\rangle }{1152}\nonumber\\
&& -\frac{ig_s G^{a}_{\alpha\beta}t^a_{ij}(\!\not\!{x}
\sigma^{\alpha\beta}+\sigma^{\alpha\beta} \!\not\!{x})}{32\pi^2x^2} +\frac{i\delta_{ij}x^2\!\not\!{x}g_s^2\langle \bar{q}\gamma_\mu t^n q\bar{q}\gamma^\mu t^n q\rangle}{3456}   -\frac{1}{8}\langle\bar{q}_j\sigma^{\mu\nu}q_i \rangle \sigma_{\mu\nu} \nonumber\\
&&   -\frac{1}{4}\langle\bar{q}_j\gamma^{\mu}q_i\rangle \gamma_{\mu }+\cdots \, ,
\end{eqnarray}
\begin{eqnarray}
S^Q_{ij}(x)&=&\frac{i}{(2\pi)^4}\int d^4k e^{-ik \cdot x} \left\{
\frac{\delta_{ij}}{\!\not\!{k}-m_Q}
-\frac{g_sG^n_{\alpha\beta}t^n_{ij}}{4}\frac{\sigma^{\alpha\beta}(\!\not\!{k}+m_Q)+(\!\not\!{k}+m_Q)
\sigma^{\alpha\beta}}{(k^2-m_Q^2)^2}\right.\nonumber\\
&& +\frac{g_s D_\alpha G^n_{\beta\lambda}t^n_{ij}(f^{\lambda\beta\alpha}+f^{\lambda\alpha\beta}) }{3(k^2-m_Q^2)^4}-\frac{g_s^2 (t^at^b)_{ij} G^a_{\alpha\beta}G^b_{\mu\nu}(f^{\alpha\beta\mu\nu}+f^{\alpha\mu\beta\nu}+f^{\alpha\mu\nu\beta}) }{4(k^2-m_Q^2)^5}\nonumber\\
&&\left.+\frac{i\langle g_s^3GGG\rangle}{48}\frac{(\!\not\!{k}+m_Q)\left[\!\not\!{k}(k^2-3m_Q^2)+2m_Q(2k^2-m_Q^2) \right](\!\not\!{k}+m_Q)}{(k^2-m_Q^2)^6}+\cdots\right\}\, ,\nonumber\\
f^{\lambda\alpha\beta}&=&(\!\not\!{k}+m_Q)\gamma^\lambda(\!\not\!{k}+m_Q)\gamma^\alpha(\!\not\!{k}+m_Q)\gamma^\beta(\!\not\!{k}+m_Q)\, ,\nonumber\\
f^{\alpha\beta\mu\nu}&=&(\!\not\!{k}+m_Q)\gamma^\alpha(\!\not\!{k}+m_Q)\gamma^\beta(\!\not\!{k}+m_Q)\gamma^\mu(\!\not\!{k}+m_Q)\gamma^\nu(\!\not\!{k}+m_Q)\, ,
\end{eqnarray}
 $t^n=\frac{\lambda^n}{2}$, the $\lambda^n$ is the Gell-Mann matrix, the $i$, $j$ are color indexes(One can consult Refs.\cite{Reinders85,Pascual1984} for the technical details in deriving the full heavy quark and light quark propagators, respectively.); then compute  the integrals both in
the coordinate and momentum spaces;  finally obtain the QCD spectral density through dispersion relation,
\begin{eqnarray}
\Pi(p^2)&=&\frac{1}{\pi}\int_{m_Q^2}^\infty \frac{{\rm Im}\Pi(s)}{s-p^2}=\int_{m_Q^2}^\infty \frac{\rho_{QCD}(s)}{s-p^2}\, .
\end{eqnarray}
In Eq.(10), we retain the terms $\langle\bar{q}_j\sigma_{\mu\nu}q_i \rangle$ and $\langle\bar{q}_j\gamma_{\mu}q_i\rangle$ originate from the Fierz re-ordering of the $\langle q_i \bar{q}_j\rangle$ to  absorb the gluons  emitted from the heavy quark lines to form $\langle\bar{q}_j g_s G^a_{\alpha\beta} t^a_{mn}\sigma_{\mu\nu} q_i \rangle$ and $\langle\bar{q}_j\gamma_{\mu}q_ig_s D_\nu G^a_{\alpha\beta}t^a_{mn}\rangle$ so as to extract the mixed condensate and four-quark condensate $\langle\bar{q}g_s\sigma G q\rangle$ and $g_s^2\langle\bar{q}q\rangle^2$, respectively. In Ref.\cite{HM-32-P}, such contributions are neglected.

\begin{figure}
 \centering
 \includegraphics[totalheight=3cm,width=14cm]{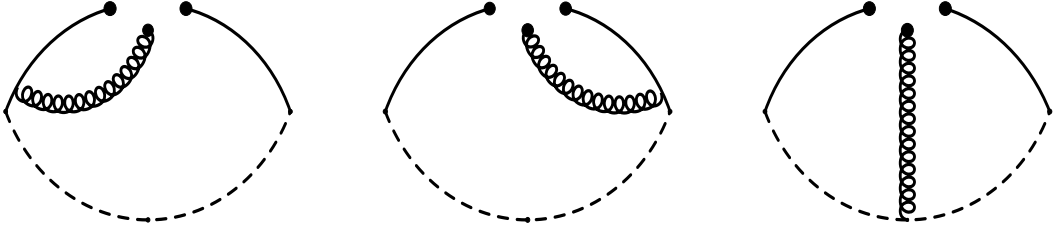}
    \caption{The diagrams contribute to the mixed condensate $\langle\bar{q}g_s \sigma G q\rangle$. }
\end{figure}

\begin{figure}
 \centering
 \includegraphics[totalheight=6cm,width=14cm]{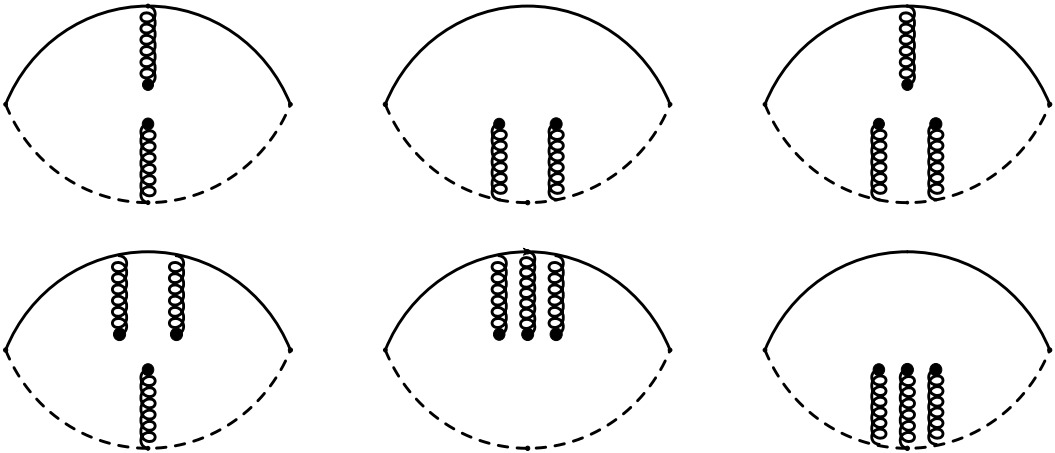}
    \caption{The diagrams contribute to the gluon condensate $\langle \frac{\alpha_sGG}{\pi}\rangle$ and three-gluon condensate $\langle g_s^3 GGG\rangle$. }
\end{figure}

\begin{figure}
 \centering
 \includegraphics[totalheight=3cm,width=14cm]{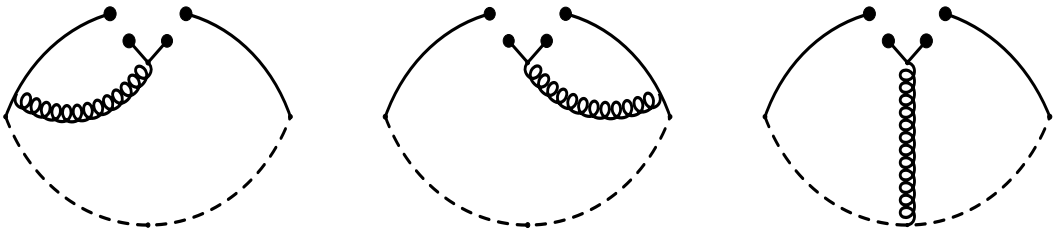}
    \caption{The diagrams contribute to the four-quark condensate $\langle\bar{q} q\rangle^2$. }
\end{figure}

We take quark-hadron duality below the continuum thresholds $s_0$ and perform the Borel transform  with respect to the variable
$P^2=-p^2$ to obtain the QCD sum rules,
\begin{eqnarray}
 f_{T}^2 M_{T}^4\exp\left(-\frac{M_{T}^2}{M^2}\right)&=& \frac{1}{10\pi^2} \int_{m_Q^2}^{s_0} ds \frac{(s-m_Q^2)^4(3s+2m_Q^2)+10m_qm_Qs(s-m_Q^2)^3}{s^3}  \exp\left(-\frac{s}{M^2}\right) \nonumber \\
  && - \frac{m_Q\langle\bar{q}g_s\sigma Gq\rangle}{2 } \exp\left(-\frac{m_Q^2}{M^2}\right)+\left(\frac{m_q\langle\bar{q}g_s\sigma Gq\rangle}{6}-\frac{2g_s^2\langle\bar{q}q\rangle^2}{81} \right)\nonumber\\
  &&\left(1+\frac{m_Q^2}{M^2} \right)\exp\left(-\frac{m_Q^2}{M^2}\right)+m_Q^2\langle \frac{\alpha_sGG}{\pi}\rangle \left[\left( \frac{1}{18} -\frac{2}{9}\frac{m_Q^2}{M^2} -\frac{5}{18}\frac{M^2}{m_Q^2}\right) \right. \nonumber\\
 && \left. \exp\left(-\frac{m_Q^2}{M^2}\right)+\left(\frac{1}{6}\frac{m_Q^2}{M^2} +\frac{2}{9}\frac{m_Q^4}{M^4}\right)\Gamma\left(0,\frac{m_Q^2}{M^2} \right) \right] +\frac{\langle g_s^3 GGG\rangle}{48\pi^2}\nonumber\\
 &&\left[\left( -\frac{20}{3} -\frac{16}{3}\frac{m_Q^2}{M^2}\right)\exp\left(-\frac{m_Q^2}{M^2}\right)+\left(12\frac{m_Q^2}{M^2}
 +\frac{16}{3}\frac{m_Q^4}{M^4}\right)\Gamma\left(0,\frac{m_Q^2}{M^2}\right)\right]\, , \nonumber\\
\end{eqnarray}
$\Gamma(0,x)=\int_0^\infty dt \frac{1}{t}e^{-xt}$.
In this article, we carry out the operator product expansion up to the vacuum condensates of dimension 6 in the leading order approximation.
In Figs.1-3, we express the contributions of the mixed condensates, four-quark condensates, gluon condensates and three-gluon condensates in terms of Feynman diagrams, which are drawn up directly from Eqs.(8-11). In the Feynman diagrams, we use the solid and dashed lines to represent the light and heavy quark propagators, respectively.
The perturbative contributions are consistent with that obtained in Ref.\cite{HM-32-P} (also that in Refs.\cite{Aliev1982,Reinders1982,Bagan1988} in the limit $m_Q\rightarrow0$), however, the contributions of  the mixed condensates differ from that obtained in Ref.\cite{HM-32-P}. In Ref.\cite{HM-32-P}, the third Feynman diagram in Fig.1 is neglected, the Feynman diagrams in Figs.2-4 are fully neglected. The contributions of the gluon condensate, three-gluon condensate and four-quark condensate  are obtained originally in this work.  As far as the contributions  of the vacuum condensates are concerned, the light quark currents and the heavy-light quark currents lead to different expressions, which do not coincide in the limit $m_Q\rightarrow0$.

If we replace the partial derivative with the covariant derivative in the interpolating currents, the following terms
\begin{eqnarray}
&&-\frac{M^2}{3}\langle\frac{\alpha_sGG}{\pi}\rangle\exp\left(- \frac{m_Q^2}{M^2}\right)-\frac{\langle g_s^3GGG\rangle}{24\pi^2}\left(\frac{5}{6}+\frac{m_Q^2}{M^2} \right)\exp\left(- \frac{m_Q^2}{M^2}\right) \nonumber\\
&&+\frac{\langle g_s^3GGG\rangle}{24\pi^2}\frac{m_Q^2}{M^2}\left(\frac{11}{6}+\frac{m_Q^2}{M^2} \right) \Gamma\left(0, \frac{m_Q^2}{M^2}\right) \, ,
\end{eqnarray}
should be added to the right side of the Eq.(13). In the leading order approximation, the perturbative terms of the QCD spectral densities are not modified, we take
into account the additional contributions by the modified vertexes,
\begin{eqnarray}
\Gamma_{\mu\nu}&=&i\left( \gamma_\mu\stackrel{\leftrightarrow}{\frac{\partial}{\partial x^\nu}} +\gamma_\nu\stackrel{\leftrightarrow}{\frac{\partial}{\partial x^\mu}}-\frac{2}{3}\widetilde{g}_{\mu\nu} \gamma^\tau\stackrel{\leftrightarrow}{\frac{\partial}{\partial x^\tau}}\right)+2g_s\left( \gamma_\mu G_\nu(x) +\gamma_\nu G_\mu(x)-\frac{2}{3}\widetilde{g}_{\mu\nu} \gamma^\tau G_\tau(x)\right)\, ,\nonumber\\
\Gamma_{\alpha\beta}&=&i\left( \gamma_\alpha\stackrel{\leftrightarrow}{\frac{\partial}{\partial y^\beta}} +\gamma_\beta\stackrel{\leftrightarrow}{\frac{\partial}{\partial y^\alpha}}-\frac{2}{3}\widetilde{g}_{\alpha\beta} \gamma^\tau\stackrel{\leftrightarrow}{\frac{\partial}{\partial y^\tau}}\right)+2g_s\left( \gamma_\alpha G_\beta(y) +\gamma_\beta G_\alpha(y)-\frac{2}{3}\widetilde{g}_{\alpha\beta} \gamma^\tau G_\tau(y)\right)\, ,\nonumber\\
\end{eqnarray}
where $G_\mu(x)=\frac{1}{2}x^\theta G_{\theta\mu}(0)+\cdots$ and $G_\alpha(y)=\frac{1}{2}y^\theta G_{\theta\alpha}(0)+\cdots=0$ in the fixed point gauge. The contributions are shown explicitly by the Feynman diagrams in Fig.4. There are no additional contributions come from the mixed quark condensate, i.e. we calculate the Feynman diagram in Fig.5 and observe  that the contribution is zero.

Differentiate   Eq.(13) with respect to  $\frac{1}{M^2}$, then eliminate the
 decay constants $f_{T}$, we obtain the QCD sum rules for
 the masses of the tensor mesons. In this article, we take into account the contributions come from the  covariant derivative.

\begin{figure}
 \centering
 \includegraphics[totalheight=6cm,width=14cm]{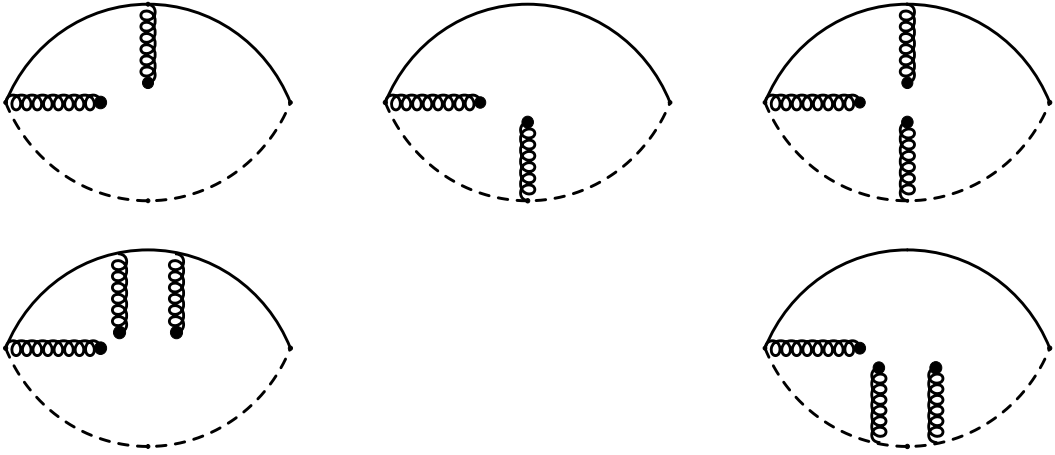}
    \caption{The additional diagrams contribute to the gluon condensate $\langle \frac{\alpha_sGG}{\pi}\rangle$ and three-gluon condensate $\langle g_s^3 GGG\rangle$ from the covariant derivative. }
\end{figure}

\begin{figure}
 \centering
 \includegraphics[totalheight=3cm,width=4cm]{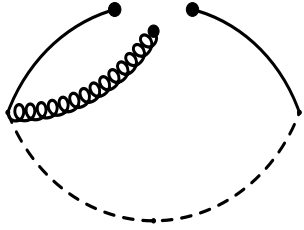}
    \caption{The additional diagram contributes  to the mixed condensate $\langle \bar{q}g_s\sigma G q\rangle$  from the covariant derivative. }
\end{figure}

\section{Numerical results and discussions}
The  masses  of the  tensor  mesons  listed in the   Review of Particle Physics are $M_{D^{*}_2(2460)^\pm}=(2464.3\pm1.6)\,\rm{MeV}$,  $M_{D^{*}_2(2460)^0}=(2461.8\pm0.7)\,\rm{MeV}$, $M_{D^{*}_{s2}(2573)}=(2571.9\pm0.8)\,\rm{MeV}$, $M_{B^{*}_2(5747)^0}=(5743\pm5)\,\rm{MeV}$,   $M_{B^{*}_{s2}(5840)^0}=(5839.96\pm0.20)\,\rm{MeV}$
\cite{PDG}.
We can take the threshold parameters  as $s^0_{D^*_2}=(8.5\pm0.5)\,\rm{GeV}^2$, $s^0_{D^*_{s2}}=(9.5\pm0.5)\,\rm{GeV}^2$, $s^0_{B^*_2}=(39\pm 1)\,\rm{GeV}^2$, $s^0_{B^*_{s2}}=(41\pm 1)\,\rm{GeV}^2$ tentatively
 to avoid the contaminations of the high resonances and continuum states, the energy gaps $\sqrt{s^0_T}-M_{T}=(0.4-0.6)\,\rm{GeV}$, the contributions of the ground states are fully included.

The quark condensates and mixed condensates  are taken to be the standard values
$\langle\bar{q}q \rangle=-(0.24\pm0.01\, \rm{GeV})^3$, $\langle\bar{s}s \rangle=(0.8\pm0.1)\langle\bar{q}q \rangle$, $\langle \bar{q}g_s \sigma G q\rangle=m_0^2\langle\bar{q}q \rangle$, $\langle \bar{s}g_s \sigma G s\rangle=m_0^2\langle\bar{s}s \rangle$, $m_0^2=(0.8\pm0.1)\,\rm{GeV}^2$  at the energy scale  $\mu=1\, \rm{GeV}$ \cite{ColangeloReview}.  The quark condensate and mixed quark condensate evolve with the   renormalization group equation,
$\langle\bar{q}q \rangle(\mu)=\langle\bar{q}q \rangle(Q)\left[\frac{\alpha_{s}(Q)}{\alpha_{s}(\mu)}\right]^{\frac{4}{9}}$,
 $\langle\bar{s}s \rangle(\mu)=\langle\bar{s}s \rangle(Q)\left[\frac{\alpha_{s}(Q)}{\alpha_{s}(\mu)}\right]^{\frac{4}{9}}$,
 $\langle\bar{q}g_s \sigma Gq \rangle(\mu)=\langle\bar{q}g_s \sigma Gq \rangle(Q)\left[\frac{\alpha_{s}(Q)}{\alpha_{s}(\mu)}\right]^{\frac{2}{27}}$
 and $\langle\bar{s}g_s \sigma Gs \rangle(\mu)=\langle\bar{s}g_s \sigma Gs \rangle(Q)\left[\frac{\alpha_{s}(Q)}{\alpha_{s}(\mu)}\right]^{\frac{2}{27}}$.
 The values of the gluon condensate and three-gluon condensate are also taken to be the standard values  $\langle \frac{\alpha_s GG}{\pi}\rangle=0.012 \,\rm{GeV}^4 $ and $\langle g_s^3 GGG\rangle=0.045\,\rm{GeV}^6$ \cite{ColangeloReview}.

In the article, we neglect the small $u$, $d$ quark masses and
take the $\overline{MS}$ masses $m_{c}(m_c)=(1.275\pm0.025)\,\rm{GeV}$, $m_{b}(m_b)=(4.18\pm0.03)\,\rm{GeV}$ and $m_s(\mu=2\,\rm{GeV})=(0.095\pm0.005)\,\rm{GeV}$
 from the Particle Data Group \cite{PDG}, and take into account
the energy-scale dependence of  the $\overline{MS}$ masses from the renormalization group equation,
\begin{eqnarray}
m_s(\mu)&=&m_s({\rm 2GeV} )\left[\frac{\alpha_{s}(\mu)}{\alpha_{s}({\rm 2GeV})}\right]^{\frac{4}{9}} \, ,\nonumber\\
m_c(\mu)&=&m_c(m_c)\left[\frac{\alpha_{s}(\mu)}{\alpha_{s}(m_c)}\right]^{\frac{12}{25}} \, ,\nonumber\\
m_b(\mu)&=&m_b(m_b)\left[\frac{\alpha_{s}(\mu)}{\alpha_{s}(m_b)}\right]^{\frac{12}{23}} \, ,\nonumber\\
\alpha_s(\mu)&=&\frac{1}{b_0t}\left[1-\frac{b_1}{b_0^2}\frac{\log t}{t} +\frac{b_1^2(\log^2{t}-\log{t}-1)+b_0b_2}{b_0^4t^2}\right]\, ,
\end{eqnarray}
  where $t=\log \frac{\mu^2}{\Lambda^2}$, $b_0=\frac{33-2n_f}{12\pi}$, $b_1=\frac{153-19n_f}{24\pi^2}$, $b_2=\frac{2857-\frac{5033}{9}n_f+\frac{325}{27}n_f^2}{128\pi^3}$,  $\Lambda=213\,\rm{MeV}$, $296\,\rm{MeV}$  and  $339\,\rm{MeV}$ for the flavors  $n_f=5$, $4$ and $3$, respectively  \cite{PDG}. In calculations, we take $n_f=4$ and  $\mu_{c(b)}=  1(3)\,\rm{GeV}$ for the charmed (bottom) tensor mesons.
 We choose the energy scales $\mu_c$ and $\mu_b$ for the charmed mesons and bottom mesons respectively  based on the crude estimation,   $\mu_c=\sqrt{M_D^2-m_c^2}\approx\sqrt{1.9^2-1.5^2}\,\rm{GeV}\approx1\,\rm{GeV}$,
  $\mu_b=\sqrt{M_B^2-m_b^2}\approx\sqrt{5.3^2-4.5^2}\,\rm{GeV}\approx3\,\rm{GeV}$, where the $M_D$ and $M_B$ are the masses of the ground states (of the pseudoscalar mesons), the $m_c$ and $m_b$ are constituent quark masses.
  The strong coupling constant $\alpha_s(\mu)$ in itself is not a physical
observable, but rather a quantity defined in the context of perturbation theory, which
enters predictions for experimentally measurable observables. We can extract the value of the $\alpha_s(\mu)$ from the experimental data at a special energy scale $\mu$, then fit the parameter $\Lambda$ with the expressions of the $\alpha_s(\mu)$ from one-loop, two-loop, three-loop, or four-loop
renormalization group equations. The values of the $\alpha_s(\mu)$ from  three-loop renormalization group equation are already compatible with that from different determination \cite{PDG}, we prefer the expression in Eq.(16), not the crude one-loop approximation.

 We impose
the two criteria (pole dominance and convergence of the operator product
expansion) on the charmed (or bottom) tensor mesons, and search for the optimal  values of the Borel parameters.
The threshold parameters, Borel parameters, pole contributions and the resulting masses and decay constants are shown explicitly in Table 1.
The pole contributions are about  $(45-80)\%$ and $(45-65)\%$ for the charmed   and bottom tensor mesons, respectively, the pole dominance is well satisfied. On the other hand, the dominant contributions come from the perturbative term, while the total contributions come from the gluon condensate and three-gluon condensates  are about  $(5-10)\%$ and $(2-5)\%$ for the  charmed    and bottom tensor  mesons, respectively, the operator product expansion is well convergent. The two criteria of the QCD sum rules are fully satisfied, we expect to obtain reasonable predictions.

In Figs.6-7, we plot the masses and decay constants with variations of the Borel parameters. From the figures, we can see that they are rather stable with variations of the Borel parameters  in the Borel windows, it is reliable to extract the masses and decay constants. The predicted masses are in excellent agreement with the  experimental data \cite{PDG}.
The predictions $f_{D_2^*}=(0.182\pm0.020)\,\rm{GeV}$, $f_{D_{s2}^*}=(0.222\pm0.021)\,\rm{GeV}$ are compatible with the values  $f_{D_2^*}=(0.225\pm0.067)\,\rm{GeV}$, $f_{D_{s2}^*}=(0.237\pm0.113)\,\rm{GeV}$ from Ref.\cite{HM-32-P}. In Ref.\cite{HM-32-P}, the contributions of the gluon condensates, three-gluon condensates, four-quark condensates and some mixed condensates  are neglected, so the present predictions are more robust.
We can take the decay constants as basic input parameters and study the revelent processes with the three-point QCD sum rules or the light-cone QCD sum rules, for example, the strong decays
\begin{eqnarray}
D_2^*(2460)^0 &\to& D^+\pi^-, \, D^{*+}\pi^-, \,D^0\pi^0, \,D^{*0}\pi^0 \, , \nonumber\\
D_{s2}^*(2573)^+&\to& D^0K^+,\, D^{*0}K^+, \,D^+K^0, \,D^{*+}K^0,\, D_s^+\pi^0,\,D_s^{*+}\pi^0\, , \nonumber\\
B_2^*(5741)^0&\to& B^+\pi^- , \, B^{*+}\pi^-, \,B^0\pi^0, \,B^{*0}\pi^0\, , \nonumber\\
B_{s2}^*(5840)^0&\to& B^+K^-, \,B^{*+}K^-,\,B^0\bar{K}^0,\,B^{*0}\bar{K}^0,\,B_s^0\pi^0,\,B_s^{*0}\pi^0\,.
\end{eqnarray}

The central values
\begin{eqnarray}
\frac{f_{D_{s2}^*}}{f_{D_{2}^*}}&=&\frac{f_{B_{s2}^*}}{f_{B_{2}^*}}=1.21 \, ,
\end{eqnarray}
the heavy quark symmetry works well. Furthermore, the $SU(3)$ breaking effects are compatible with the experimental data \cite{Rosner},
\begin{eqnarray}
\frac{f_{D_{s}}}{f_{D}}&=&1.258 \pm 0.038\, ,
\end{eqnarray}
the approximation $\frac{f_{D_{s2}^*}}{f_{D_{2}^*}}\approx\frac{f_{B_{s2}^*}}{f_{B_{2}^*}}\approx\frac{f_{D_{s}}}{f_{D}}$ is reasonable.

\begin{table}
\begin{center}
\begin{tabular}{|c|c|c|c|c|c|c|c|}\hline\hline
                    & $M^2 (\rm{GeV}^2)$  & $s_0 (\rm{GeV}^2)$   & pole         & $M_T(\rm{GeV})$    & $f_{T}(\rm{MeV})$   \\ \hline
  $D_2^*(2460)$     & $1.5-2.1$           & $8.5\pm0.5$          & $(44-78)\%$  & $2.46\pm0.09$      & $182\pm20$          \\ \hline
 $D_{s2}^*(2573)$   & $1.6-2.4$           & $9.5\pm0.5$          & $(48-83)\%$  & $2.58\pm0.09$      & $222\pm21$          \\ \hline
   $B_2^*(5747)$    & $4.6-5.4$           & $39\pm1$             & $(44-65)\%$  & $5.73\pm0.06$      & $110\pm11$          \\ \hline
 $B_{s2}^*(5840)$   & $5.2-6.0$           & $41\pm1$             & $(46-64)\%$  & $5.84\pm0.06$      & $134\pm11$          \\ \hline
   \hline
\end{tabular}
\end{center}
\caption{ The Borel parameters, continuum threshold parameters, pole contributions, masses and decay constants for the heavy tensor  mesons. }
\end{table}

\begin{figure}
\centering
\includegraphics[totalheight=6cm,width=7cm]{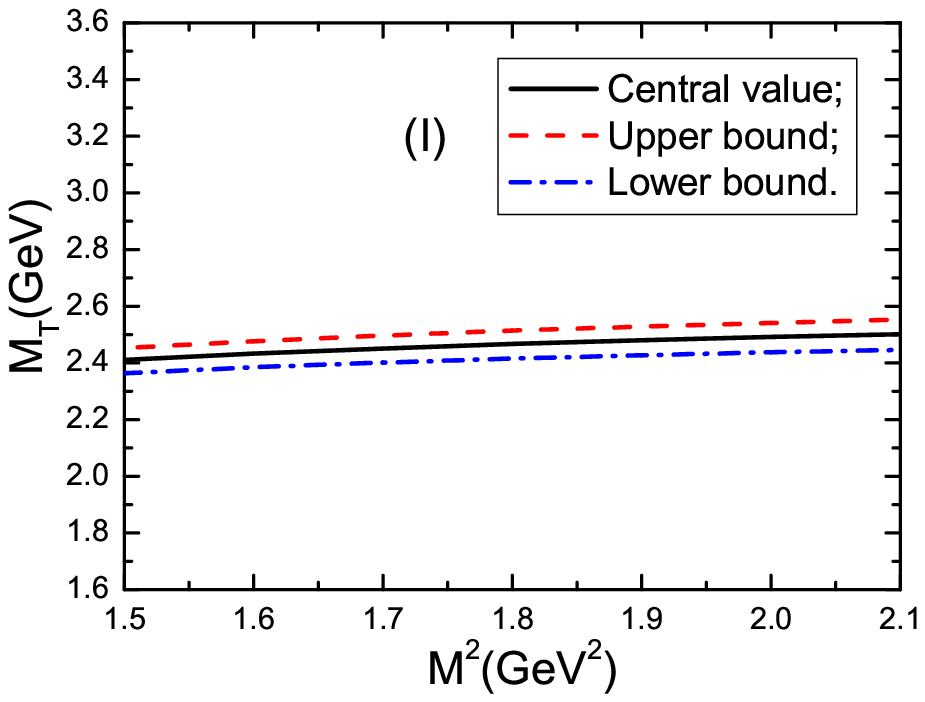}
\includegraphics[totalheight=6cm,width=7cm]{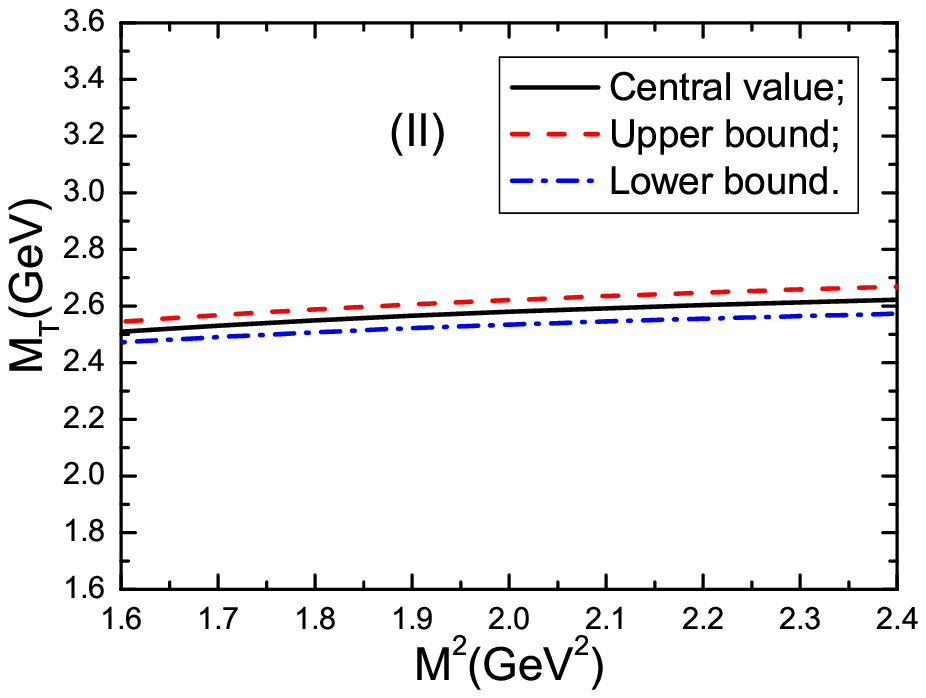}
\includegraphics[totalheight=6cm,width=7cm]{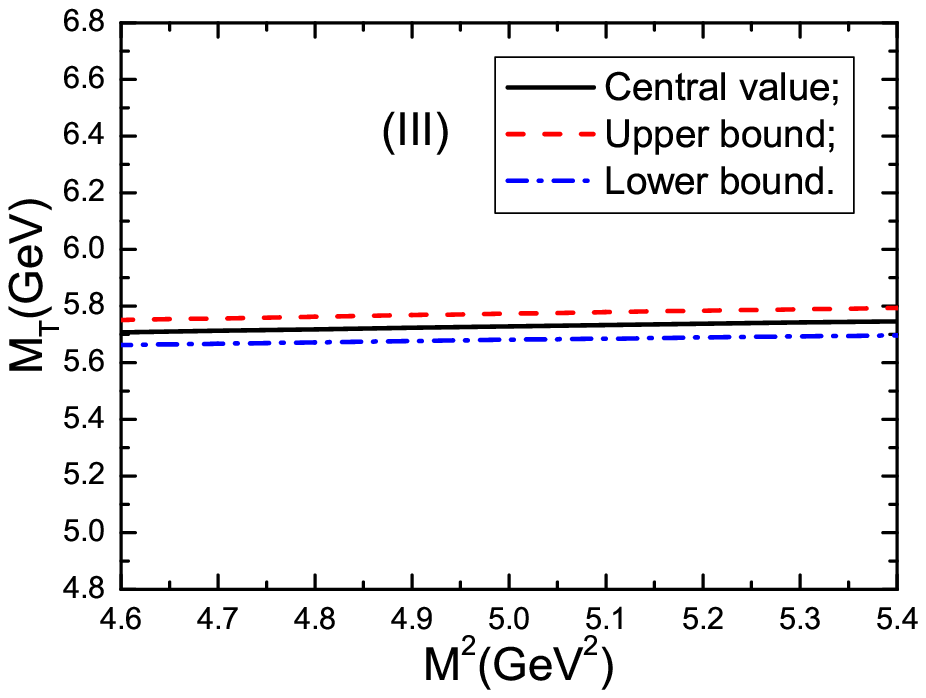}
\includegraphics[totalheight=6cm,width=7cm]{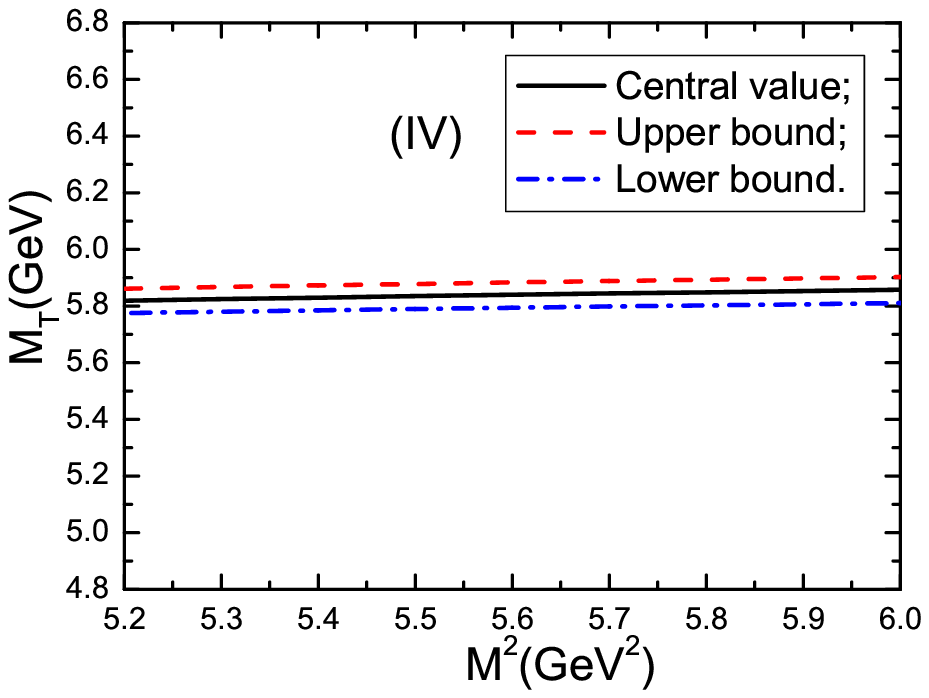}
  \caption{ The masses of the tensor mesons with variations of the  Borel parameters $M^2$, where the (I), (II), (III), (IV) denote the $D_2^*(2460)$, $D_{s2}^*(2573)$, $B_2^*(5747)$, $B_{s2}^*(5840)$, respectively. }
\end{figure}

\begin{figure}
\centering
\includegraphics[totalheight=6cm,width=7cm]{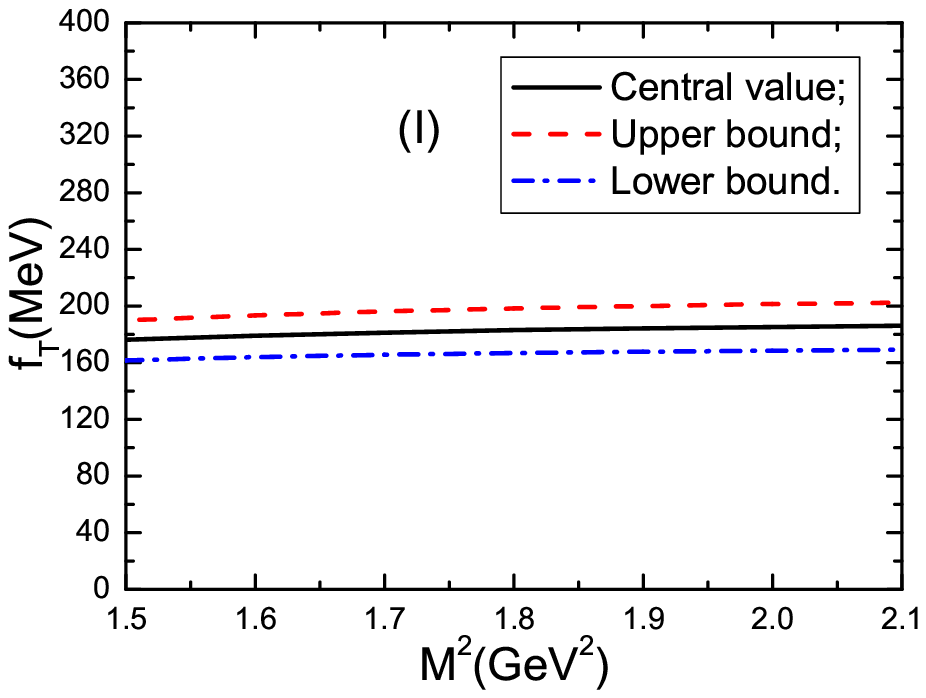}
\includegraphics[totalheight=6cm,width=7cm]{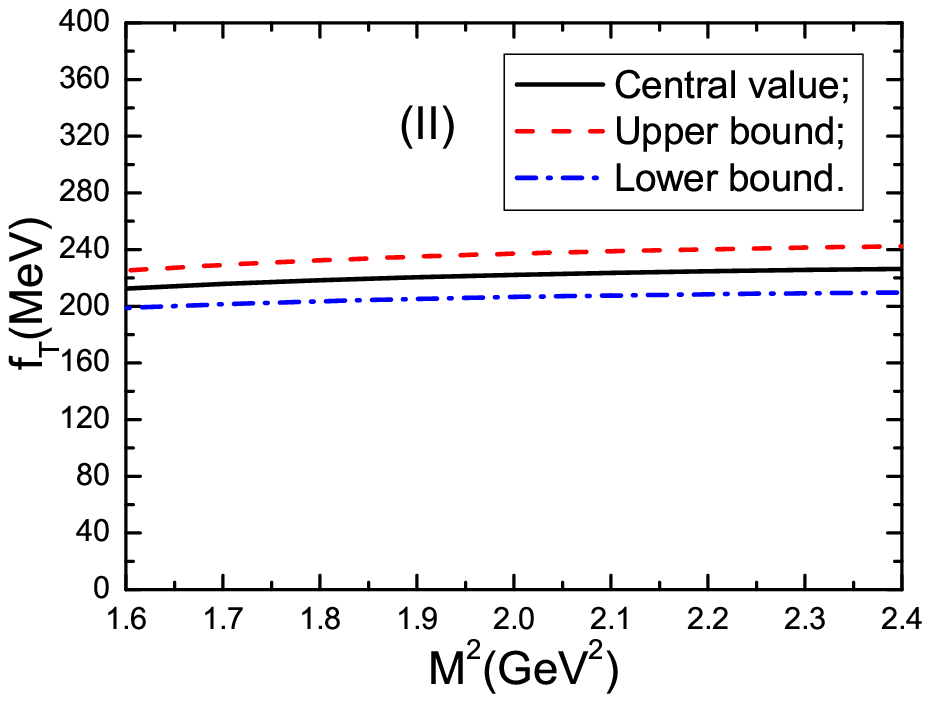}
\includegraphics[totalheight=6cm,width=7cm]{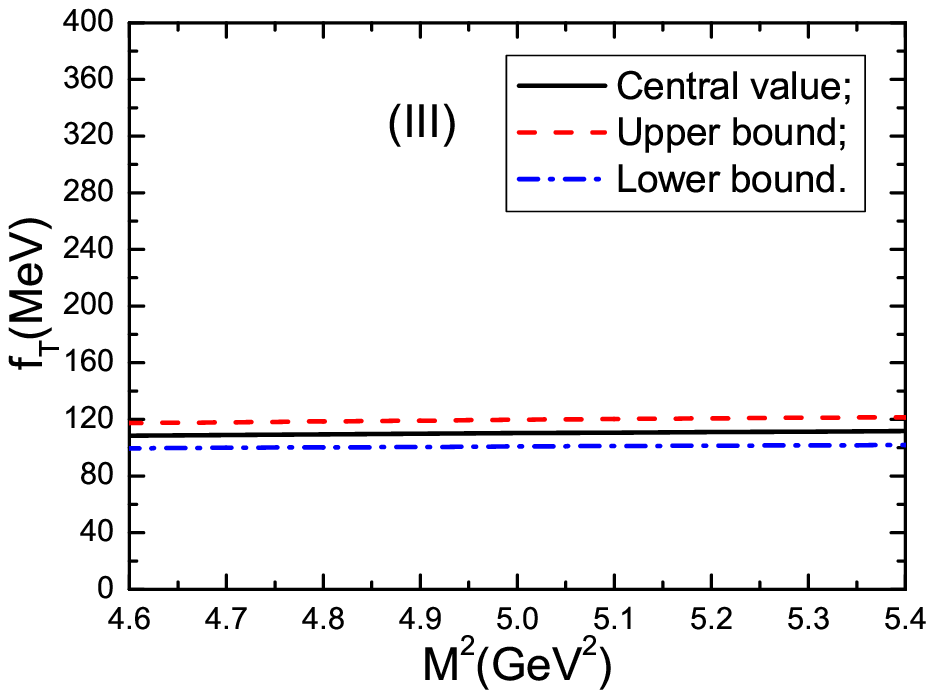}
\includegraphics[totalheight=6cm,width=7cm]{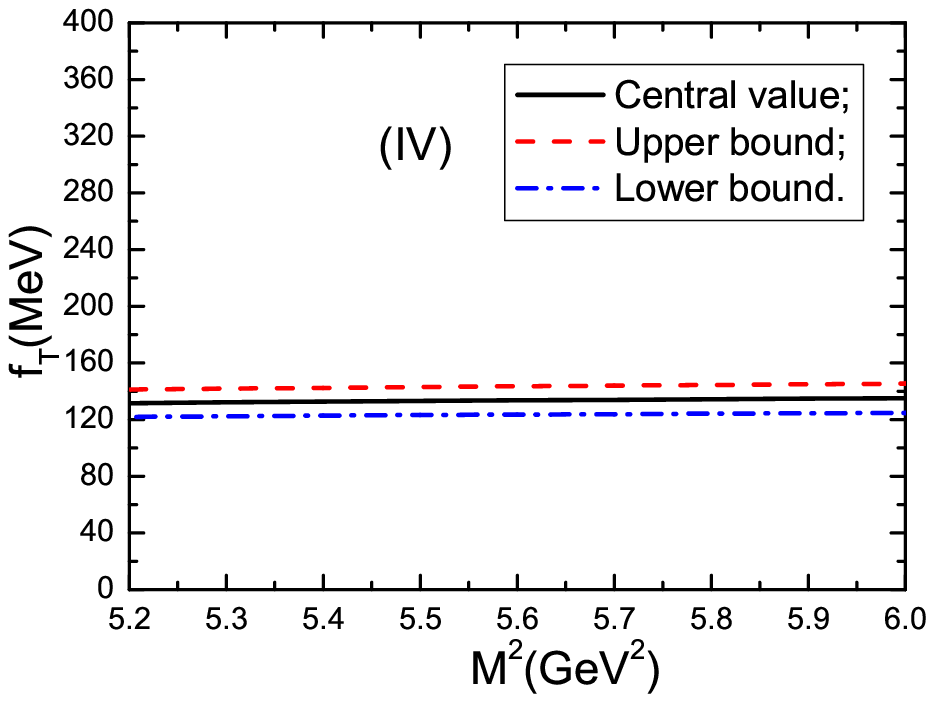}
  \caption{ The decay constants  of the tensor mesons with variations of the  Borel parameters $M^2$, where the (I), (II), (III), (IV) denote the $D_2^*(2460)$, $D_{s2}^*(2573)$, $B_2^*(5747)$, $B_{s2}^*(5840)$, respectively. }
\end{figure}

If we use the non-covariant currents instead of the covariant currents, the gluon condensate and the three-gluon condensate in Eq.(14) have no contributions,
the masses and the decay constants change as
\begin{eqnarray}
\delta M_{D_2^*}&=&-27\,{\rm{MeV}}\, , \,\,\, \,\,\,\delta f_{D_2^*}=3\,\rm{MeV}\, , \nonumber\\
\delta M_{D_{s2}^*}&=&-20\,{\rm{MeV}}\, , \,\,\, \,\,\,\delta f_{D_{s2}^*}=2\,\rm{MeV}\, , \nonumber\\
\delta M_{B_2^*}&=&-14\,{\rm{MeV}}\, , \,\,\,\,\,\, \delta f_{B_2^*}\approx 0\,\rm{MeV}\, , \nonumber\\
\delta M_{B_{s2}^*}&=&-10\,{\rm{MeV}}\, , \,\,\, \,\,\,\delta f_{B_{s2}^*}\approx 0\,\rm{MeV}\, .
\end{eqnarray}

There are effective cancelations among the contributions of the three-gluon condensate $\langle g_s^3GGG\rangle$ from different Feynman diagrams, and among the contributions of the three-gluon condensate $\langle g_s^3GGG\rangle$ and the four quark condensate  $g_s^2\langle\bar{q}q\rangle^2$. The masses and the decay constants remain almost unchanged if  the dimension-6 vacuum condensates are neglected,
\begin{eqnarray}
\delta M_{D_2^*}&\approx& 0\,{\rm{MeV}}\, , \,\,\, \,\,\,\delta f_{D_2^*}\approx 0\,\rm{MeV}\, , \nonumber\\
\delta M_{D_{s2}^*}&\approx& 0\,{\rm{MeV}}\, , \,\,\, \,\,\,\delta f_{D_{s2}^*}\approx 0\,\rm{MeV}\, , \nonumber\\
\delta M_{B_2^*}&\approx&  0\,{\rm{MeV}}\, , \,\,\, \,\,\,\delta f_{B_2^*}\approx 0\,\rm{MeV}\, , \nonumber\\
\delta M_{B_{s2}^*}&\approx&  0\,{\rm{MeV}}\, , \,\,\,\,\,\, \delta f_{B_{s2}^*}\approx 0\,\rm{MeV}\, .
\end{eqnarray}

In this article, we neglect the perturbative $\alpha_s$ corrections. In the massless limit, taking into accounting the perturbative  $\alpha_s$ corrections amounts  to  multiplying  the perturbative terms by a factor
 $\left(1-\frac{\alpha_s}{\pi}\right)$ \cite{Reinders1982}.  Now, we estimate the  perturbative $\alpha_s$ contributions by multiplying the perturbative terms by the factor  $\left(1-\frac{\alpha_s}{\pi}\right)$, which leads  to the following changes,
\begin{eqnarray}
\delta M_{D_2^*}&\approx& 0\,{\rm{MeV}}\, , \,\,\,\,\,\,  \delta f_{D_2^*}=-15\,\rm{MeV}\, , \nonumber\\
\delta M_{D_{s2}^*}&\approx& 0\,{\rm{MeV}}\, , \,\,\,\,\,\, \delta f_{D_{s2}^*}=-18\,\rm{MeV}\, , \nonumber\\
\delta M_{B_2^*}&\approx& 0\,{\rm{MeV}}\, , \,\,\, \,\,\,\delta f_{B_2^*}=-5\,\rm{MeV}\, , \nonumber\\
\delta M_{B_{s2}^*}&\approx& 0\,{\rm{MeV}}\, , \,\,\,\,\,\, \delta f_{B_{s2}^*}=-6\,\rm{MeV}\, .
\end{eqnarray}

In calculations, we observe that   the masses decrease monotonously
with increase of the energy scales while the decay constants  increase  monotonously
with increase of the energy scales. If we enlarge the energy scales by
  $ \mu_{c(b)} \to  \mu_{c(b)}+ 300\,\rm{MeV}$, then
\begin{eqnarray}
\delta M_{D_2^*}&=&-44\,{\rm{MeV}}\, , \,\,\, \,\,\,\delta f_{D_2^*}=27\,\rm{MeV}\, , \nonumber\\
\delta M_{D_{s2}^*}&=&-46\,{\rm{MeV}}\, , \,\,\, \,\,\,\delta f_{D_{s2}^*}=27\,\rm{MeV}\, , \nonumber\\
\delta M_{B_2^*}&=&-37\,{\rm{MeV}}\, , \,\,\,\,\,\, \delta f_{B_2^*}=11\,\rm{MeV}\, , \nonumber\\
\delta M_{B_{s2}^*}&=&-36\,{\rm{MeV}}\, , \,\,\, \,\,\,\delta f_{B_{s2}^*}=12\,\rm{MeV}\, ,
\end{eqnarray}
the changes are sizeable, but they are small compared to the energy scale augment $300\,\rm{MeV}$.
The correlation functions $\Pi(p^2)$ can be written as
\begin{eqnarray}
\Pi(p^2)&=&\int_{ m^2_Q(\mu)}^{s_0} ds \frac{\rho_{QCD}(s,\mu)}{s-p^2}+\int_{s_0}^\infty ds \frac{\rho_{QCD}(s,\mu)}{s-p^2} \, ,
\end{eqnarray}
through dispersion relation at the QCD side, and they are scale independent,
$\frac{d}{d\mu}\Pi(p^2)=0$,
which does not amount to
\begin{eqnarray}
\frac{d}{d\mu}\int_{m^2_Q(\mu)}^{s_0} ds \frac{\rho_{QCD}(s,\mu)}{s-p^2}\rightarrow 0 \, ,
\end{eqnarray}
  as the perturbative corrections to all orders are neglected and truncations $s_0$ set in. The correlation between the threshold $m^2_Q(\mu)$ and continuum threshold $s_0$ is unknown.
We cannot obtain energy scale independent QCD sum rules, but we can choose the reasonable   energy scales based on some theoretical analysis.

\section{Conclusion}
In this article, we calculate the contributions of the vacuum condensates up to dimension-6 in the operator product expansion,  study the masses and decay constants of the heavy tensor mesons using the QCD sum rules.
The predicted masses of the $D_2^*(2460)$, $D_{s2}^*(2573)$, $B_2^*(5747)$, $B_{s2}^*(5840)$ are in excellent agreement with the experimental data, while the ratios of the decay constants $\frac{f_{D_{s2}^*}}{f_{D_{2}^*}}\approx\frac{f_{B_{s2}^*}}{f_{B_{2}^*}}\approx\frac{f_{D_{s}}}{f_{D}}\mid_{\rm exp}$, where the exp denotes the experimental value. The decay constants can be taken as basic input parameters in studying the strong decays with the three-point QCD sum rules or the light-cone QCD sum rules.

\section*{Acknowledgements}
This  work is supported by National Natural Science Foundation,
Grant Numbers 11375063, the Fundamental Research Funds for the
Central Universities,  and Natural Science Foundation of Hebei province, Grant Number A2014502017.

\end{document}